\documentclass[letterpaper,twocolumn,preprintnumbers,secnumarabic,amsmath,amssymb,superscriptaddress]{revtex4-1}
\usepackage{amssymb,amsmath}
\usepackage{ulem} %for strikethrough
\usepackage{bm} %math bold symbols
\usepackage{booktabs} %some alignment stuff
\usepackage{array}
\usepackage{latexsym}
\usepackage{graphicx}
\usepackage{color}
\usepackage{datetime}

\usepackage{xcolor}  
\usepackage{color, colortbl}
\usepackage{multirow,tabularx,ragged2e}
\usepackage{physics}

\usepackage{verbatim}
\usepackage{chngpage} % allows for temporary adjustment of side margins

\usepackage{psfrag}

\usepackage{mciteplus}

\usepackage[colorlinks=true,      linkcolor=blue,      urlcolor=blue,      
            filecolor=blue,      citecolor=blue,       pdfstartview=FitH,     
						pdfpagemode=UseNone,      bookmarksopen=true]{hyperref}  % LINKS
\usepackage[all]{hypcap}     %should be loaded AFTER hyperref, which should otherwise be loaded last.s

\definecolor{cardinal}{rgb}{0.6,0,0}
\definecolor{darkgreen}{rgb}{0,0.45,0}
\definecolor{golden}{rgb}{0.92, 0.7, 0}
\definecolor{midnight}{rgb}{0, 0, 0.5}
\definecolor{darkblue}{rgb}{0.2, 0, 0.8}
\definecolor{brown}{rgb}{0.5, 0.3, 0}
\definecolor{purple}{rgb}{0.5, 0, 0.5}

\def\newstratum{hyperstratum}
\def\newstrata{hyperstrata}

\def\IC{\mathbb{C}}

\def\coeff#1#2{{\textstyle \frac{#1}{#2}}}

\def\IR{\mathbb{R}}
\def\IT{\mathbb{T}}

\def\cQ{{\cal Q}}

\begin{document}
%%%%%%%%%%%%%%%%%%%%%%%%%%%%%%

\title{Themelia: the irreducible microstructure of black holes}

 \author{Iosif Bena}
\affiliation{Institut de Physique Th\'eorique,  Universit\'e Paris Saclay,  CEA, CNRS, F-91191 Gif-sur-Yvette, France}

\author{Nejc  \v{C}eplak } 
\affiliation{Institut de Physique Th\'eorique,  Universit\'e Paris Saclay,  CEA, CNRS, F-91191 Gif-sur-Yvette, France}

\author{Shaun D. Hampton}
\affiliation{Institut de Physique Th\'eorique,  Universit\'e Paris Saclay,  CEA, CNRS, F-91191 Gif-sur-Yvette, France}

\author{Anthony Houppe} 
\affiliation{Institut de Physique Th\'eorique,  Universit\'e Paris Saclay,  CEA, CNRS, F-91191 Gif-sur-Yvette, France}
 
\author{Dimitrios Toulikas} 
\affiliation{Institut de Physique Th\'eorique,  Universit\'e Paris Saclay,  CEA, CNRS, F-91191 Gif-sur-Yvette, France}

\author{Nicholas P. Warner}
\affiliation{Institut de Physique Th\'eorique,  Universit\'e Paris Saclay,  CEA, CNRS, F-91191 Gif-sur-Yvette, France}
\affiliation{Department of Physics and Astronomy and Department of Mathematics, University of Southern California, Los Angeles, CA 90089, USA\\ 
\vspace*{3mm}{\rm \textsf{ iosif.bena@ipht.fr,  nejc.ceplak@ipht.fr, shaun.hampton@ipht.fr, anthony.houppe@ipht.fr, dimitrios.toulikas@ipht.fr, warner@usc.edu} }}

\begin{abstract}
\noindent We argue that the fundamental ``atomic objects'' in string theory are {\it themelia}: extended objects that have 16 supersymmetries {\it locally}. We show that all existing smooth horizonless microstate geometries can be seen as bound states of themelia, and we conjecture that all such bound states with suitable KKM charges will give rise to microstate geometries. We also construct the most general themelion with a three-torus isometry and show that it interpolates between superstrata and the super-maze.

\end{abstract}

\maketitle

\vspace{-1mm}

Whether you describe it in General Relativity or think of it as a strongly-coupled quantum object, a black hole must necessarily reduce matter to its most fundamental constituents.   From the perspective of string theory, this is usually interpreted as meaning strings and branes, but we will argue that the full range of fundamental constituents must  include   more general species of objects that we will call {\it themelia}.   A themelion is defined to be any object in string theory that {\it locally preserves} $16$ supercharges.   This certainly includes fundamental strings and branes, but a themelion can carry multiple charges and preserve less supersymmetry when taken as a whole.  A  themelion will typically have varying charge densities along a non-trivial profile, but the defining idea is that when one ``zooms in'' on a small segment of the themelion, the localized part  preserves sixteen supersymmetries and those supersymmetries will generically depend on their location on the themelion.

Our purpose here is not only to characterize some large families of themelia, but also to show that they play the central role in the description of black-hole microstructure, and that they are necessarily the irreducible constituents of a supersymmetric fuzzball.  As we will describe,  themelia not only include all known supersymmetric microstate geometries, but also  greatly extend their range. Indeed, a central result of this paper will be to exhibit  themelia that embed the  microstate geometries known as superstrata \cite{Bena:2015bea,Bena:2016ypk,Bena:2017xbt}  into highly fractionated brane configurations that include the super-maze \cite{Bena:2022wpl}.

The fuzzball paradigm  \cite{Mathur:2009hf,Bena:2022rna} seeks a gravitational and quantum description of black holes,  and their microstructure, in terms of  horizonless objects in string theory.  The idea is that, because individual microstates have no entropy, they cannot have a horizon, and that horizons only arise through ensemble averaging.  Fuzzballs are supposed to represent a new phase  that emerges when matter is compressed to black-hole  densities, and this new phase prevents the formation of a horizon or singularity.  

The challenge has been to formulate fuzzballs more precisely  \cite{Bena:2013dka} and this has been done largely through the construction of huge families of examples: In particular, microstate geometries are realizations of the fuzzball paradigm in terms of smooth solutions to supergravity.  What is perhaps most startling about this program is the  extent to which it can be realized. (For  recent reviews, see \cite{Bena:2022rna,Bena:2022ldq}.)  This extensive body of work has also led to a much deeper understanding of the new phase of matter that underlies fuzzballs, and hence our proposal that their fundamental constituents should be themelia.

Fundamental constituents must  themselves be horizonless.  But this is not sufficient:   there are  ``horizonless'' string configurations, like the unadorned D1-D5 solution,  that still have microstructure.  While the classical horizon of such an object has vanishing area, it can be argued to have a ``Planck-scale horizon'' that accounts for the entropy of its microstructure \cite{Sen:1995in}.   On the other hand, the sixteen supersymmetries of the themelion not only preclude it from having a horizon, even at the Planck scale, but also makes it a fundamental bound state, an indivisible ``atomic object'' of string theory -- hence  the building block of fuzzballs. 

Objects with 16 supersymmetries  can always be dualized to a system consisting of a single species of brane, such as a  stack of F1's, or the empty space of a KKM.  A generic themelion is, however, highly non-trivial: It only has 16 supersymmetries  {\it locally}, and so the ``trivializing'' duality transformation depends on the location on the themelion profile.
In any fixed duality frame, a themelion can carry a huge range of charges that {\it vary} with location.   Some of these will average to zero over the themelion, and some will average to non-zero values.  We will refer to these as {\it dipolar}  and {\it global} charges respectively.  The global charges determine the overall  supersymmetry preserved by the themelion. 

The important point about themelia is that, individually, each one has essentially no microstructure and, upon dualizations, can be characterized locally using string theory (as an F1)  or geometry (as a KKM).  However, globally, themelia can have huge moduli spaces, expressed in terms of shapes and charge densities, and so they can encode a vast number of microstates  within  their  configuration spaces.   Conversely, because of the 16 supersymmetries, the moduli space of a themelion cannot {\it ipso facto} contain any black holes, or give rise to horizons. 

Superstrata were originally conjectured to exist \cite{Bena:2011uw} entirely based on the underlying principle of themelia: namely, $16$  supersymmetries locally.  Five years later, large families of superstratum supergravity solutions were explicitly constructed \cite{Bena:2015bea,Bena:2016ypk,Bena:2017xbt} but their connection with the themelion of  \cite{Bena:2011uw} was not clear. 

In this Letter we exhibit the themelion structure of the superstratum solution, and show {that the supergravity constraints imposed by smoothness} are equivalent to requiring that themelia have 16 local supercharges.
 We also reveal a similar structure in the recently constructed  ``vector superstrata'' \cite{Ceplak:2022}.

In addition to superstrata, one should recall that there is another huge family of smooth horizonless microstate geometries: the bubbling solutions constructed from ambi-polar Gibbons-Hawking (GH) geometries \cite{Bena:2005va, Berglund:2005vb, Bena:2007kg}. Upon reduction to 10 dimensions along the GH fiber, one obtains a multi-center solution \cite{Bates:2003vx} where each center has 16 supercharges \cite{Balasubramanian:2006gi} (and hence is a themelion).  

Based on our observations, we conjecture a relation between bound states of themelia and smooth horizonless supergravity solutions:

{\bf The Themelion Conjecture:}\\
{\it All smooth horizonless solutions come from bound states of  themelia with KKM charge. All bound states of such themelia give rise to a smooth supergravity solution.}

As we remarked above, each themelion can have a large moduli space. Furthermore, combining different themelia can lead to an enhancement of the moduli space. As we will show, the superstratum solution is a combination of two themelia, each of which involves a function of one variable and yet, the generic superstratum solution is expected to be parameterized by arbitrary functions of three variables \cite{Heidmann:2019zws}.  Therefore, our conjecture does not imply that a choice of component themelia leads to a unique smooth horizonless solution.

\vspace{-5mm}
%%%%%%%%%%%%%%%%%%%%%%%%%%%%%%%%%%%%%
\section{Themelia}
\label{Sect:Projectors}
%%%%%%%%%%%%%%%%%%%%%%%%%%%%%%%%%%%%%
\vspace{-3mm}

To pin down the structure of a particular themelion, one must first specify its global charges, and the amount of supersymmetry it will preserve overall.  One then chooses dipolar charges as ``glue'' that will bind the global object into a bound state with 16 supersymmetries locally \cite{Bena:2022wpl}.  There are typically multiple choices for such glue and, as we will see, one can often combine different types of glue to create an even larger themelion moduli space.  The choice of these dipole charges is also motivated by the underlying physics of the themelion.

The construction, of course, depends on the duality frame and the charges we want the themelion to carry. Here we will work exclusively in the Type IIA/M-theory frame and focus on themelia that have the F1, NS5 and P charges of a supersymmetric black hole. We can uplift everything to M-theory, where the supersymmetries, $\cQ$,  are 32-component spinors and the themelion building blocks are M2 and M5 branes, momentum, $P$, and KKM charge.  
The supersymmetries of the themelion are then defined by projectors involving gamma matrices.
\begin{equation}
 \Pi \cQ ~=~ 0,  \qquad \Pi= \coeff{1}{2} \,(1 ~+~ P),
 \label{susyproj}
\end{equation}
where the matrices, $P$,  are given by \cite{Bena:2011uw}:
\begin{align}
&P_{\rm M2 (12) } = \Gamma^{012}  \,, \qquad  P_{\rm M5 (12345)} = \Gamma^{012345}\notag\\
&P_{\rm P(1) } = \Gamma^{01}  \,,   \quad P_{\rm KKM(123456;7)} = \Gamma^{0123456} = \Gamma^{789\,10} \,.
\nonumber
% \label{MProjectors}
\end{align}
Here the indices indicate the directions along which the branes extend. For the KKM the last entry denotes the ``special direction'' of the fibration.

We will also focus on the  $\IT^4$ compactification of IIA supergravity to six dimensions and will denote the M-theory circle by $z$ and the torus directions as $1,2,3,4$.  We will also introduce two other circles: an $S^1 (y)$ corresponding to the common direction of the F1 strings and NS5 branes that give global charges,
%compactification to five dimensions, 
and a ``space-time''  $\psi$-circle transverse to the $\IT^4 \times S^1 (y)$.  If the $\psi$ direction is non-compact then charges corresponding to branes wrapping this direction %and to momentum along this direction 
are necessarily dipolar.   
We will also find it useful to use the standard  IIA nomenclature:  NS5, F1, D2 , D4 and D6 to encode the way in which the M-theory objects wrap the $z$-circle.

An archetypical example of the relation between themelia and supergravity solution comes from the Lunin-Mathur 8-supercharge solution \cite{Lunin:2001fv}. This is a smooth horizonless solution, which can be thought of as a supertube \cite{Mateos:2001qs} with  F1$(y)$ and  NS5$(1234y)$ charges and a KKM$(1234\psi z;y)$ dipole charge.   The global supersymmetries are defined by the  vectors, $\xi$, satisfying: 
\begin{equation}
\coeff {1}{ 2}(1 ~+~ \Gamma^{0yz} ) \, \xi  ~=~ 0\,, \quad   \coeff {1}{ 2} (1 ~+~  \Gamma^{01234 y} ) \, \xi  ~=~ 0 \,.
\label{susies1}
\end{equation}

This solution is made of two themelia. The first themelion is at the ``supertube locus'' and has F1 and NS5 charges, as well as a glue coming from the angular momentum P$(\psi)$ along the supertube direction, $\psi$, and the KKM$(1234\psi z;y)$. Its projector has  \cite{Bena:2011uw}:
\begin{equation}
P~=~ \lambda_1 \Gamma^{0yz} ~+~ \lambda_2 \Gamma^{01234 y}  ~+~\lambda_3 \Gamma^{0\psi} ~+~ \lambda_4  \Gamma^{01234\psi z} \,.
\label{supertube}
\end{equation}
If one takes $\lambda_1 = \cos^2\phi$, $\lambda_2 = \sin^2\phi$ and $\lambda_3 =  -\lambda_4 =  \sin\phi \cos \phi$, then one has: 
\begin{equation}
P^2 ~=~1 \,,  \qquad \Pi \, \xi  ~=~  \coeff {1}{ 2} (1+P) \, \xi  ~=~ 0\,.
\label{Pident}
\end{equation}
The identity $\Pi^2 = \Pi$ means that $\Pi$ only has eigenvalues $0$ or $1$ and, combined with the fact that the products of gamma matrices are traceless, it means that $\Pi$ must have sixteen null vectors, preserving 16 supersymmetries. However eight of those supersymmetries depend upon the parameter, $\phi$, and eight are independent of $\phi$ and are determined by the global F1 and NS5 charges as in (\ref{susies1}).  

The second themelion is a bit less obvious when the supertube is in  $\IR^4$, but can be seen easily if one embeds the Lunin-Mathur geometry in Taub-NUT \cite{Bena:2005ay, Bena:2005ni}: The ``center-of-space'' themelion is a KKM wrapping the $(y1234z)$ directions and with a special direction $\psi$. This reflects a more general principle: to reveal the themelion sources of a solution one should write it as a fibration over a spatial $\IR^3$. The themelia are located where the fibers degenerate.

Supergravity superstrata of \cite{Bena:2015bea,Bena:2016ypk,Bena:2017xbt} are built by adding momentum waves along $y$ to a generalized supertube solution. These momentum waves cannot be sourced on the themelion at the supertube locus, where the $y$-circle degenerates. Instead, they are sourced at the center of space, promoting the simple KKM themelion to a much more complicated momentum-carrying themelion.

To recast the original supergravity superstrata \cite{Bena:2015bea,Bena:2016ypk,Bena:2017xbt} in our IIA/M-theory frame, we perform an S-duality and a T-duality along one direction of the $\IT^4$, which we can choose to be $x_1$. By writing the superstratum solution as circle fibrations over an $\IR^3$ base, one can easily read off all the themelion charges from the singularities in the fluxes and metric functions.  In particular, there are warp factors that diverge at the supertube location. These correspond to the global F1 (M2$(yz)$) and NS5 (M5$(y1234)$) charges, and equal amounts of dipolar M2$(1y)$ and M5$(z234y)$ charges. At the center of space there are also singularities corresponding to the same set of branes wrapped on the $\psi$ direction:  M2$(\psi z)$, M5$(\psi 1234)$ and equal amounts of M2$(\psi 1)$ and M5$(\psi 234z)$. These branes are dipolar. There is also the KKM with special direction $\psi$ at the center of space. 

In this Letter, we will focus on three-charge themelia (with four global supercharges) carrying the charges F1$(y)$, NS5$(1234y)$ and P$(y)$ in Type IIA/M-theory.  The  global supersymmetries are given by:
\begin{equation}
\begin{aligned}
&\coeff {1}{ 2}(1 ~+~ \Gamma^{0yz} ) \, \xi  ~=~ 0\,, \quad   \coeff {1}{ 2} (1 ~+~  \Gamma^{01234 y} ) \, \xi  ~=~ 0\,, \\
  & \coeff {1}{ 2} (1 ~+~ \Gamma^{0y} ) \, \xi  ~=~ 0 \,.
 \end{aligned}
\label{susies2}
\end{equation}
The most general themelion projector with these charges and the dipole charges corresponding to the superstrata considered above is:
\begin{equation}
\begin{aligned}
P &=(\alpha_1 \Gamma^{0yz}   + \alpha_2   \Gamma^{0 y1234} +  \alpha_3 \Gamma^{0y} + \alpha_{4}    \Gamma^{0y 1234 z}) \\
 & \ + (\alpha_5 \Gamma^{0\psi z}  + \alpha_{6} \Gamma^{0\psi1234 }   + \alpha_{7}   \Gamma^{0 \psi}  + \alpha_8  \Gamma^{0\psi 1234  z})   \\
&\ +  (\alpha_9 \Gamma^{0y1 }   +  \alpha_{10} \Gamma^{0y 234 z})  + ( \alpha_{11} \Gamma^{0 \psi1 }  +  \alpha_{12} \Gamma^{0 \psi 234  z} ) \,,
\end{aligned}
\label{Pss1}
\end{equation}
where the $\alpha_j$ can be interpreted as local charge densities  divided by the mass density. 

The global supercharge condition (\ref{susies2}) leads to  the following linear constraints: 
\begin{eqnarray}
%\begin{aligned}
 &\alpha_1 + \alpha_2 +\alpha_3+\alpha_4=1 \,, \label{mass} \\
 & \alpha_5 +\alpha_6 + \alpha_7  +\alpha_8  = 0\,, \label{bubble}\\
 &\alpha_{10}=-\alpha_{9}~,~~~\alpha_{11}=-\alpha_{12} \,,
  \label{z4charges} 
%     \end{aligned}
\end{eqnarray}
and the projector condition $P^2 =1$ in (\ref{Pident}), leads to several quadratic conditions, which include, for example:
\begin{equation}
  (  \alpha_1 \alpha_2 + \alpha_3 \alpha_4 - \alpha_9^2) + (\alpha_5 \alpha_6 + \alpha_7 \alpha_8 - \alpha_{11}^2) = 0 \,.
\label{quadratic1}
\end{equation}
 The complete solution to the {\it themelion constraints}, obtained using \eqref{Pss1} in equations  (\ref{Pident}) and  (\ref{susies2}) 
 is given by  (\ref{reparam1}) and (\ref{uvwxyz}) with  $\theta_2 = \frac{\pi}{2}$, $\varphi_2 =0$. 

Our first result  is that: {\it All existing microstate geometries - both  bubbling solutions and  superstrata - are bound states of multiple themelia defined by \eqref{Pss1}.}

We find that all the themelion constraints have counterparts in the supergravity solutions. However, the relation is subtle. First, the themelion analysis is done for a non-backreacted brane probe in flat space. Thus, to link the supergravity charges to the $\alpha_i$ we have to arrange that themelion's environment be that of  empty space: in particular one must perform large gauge transformations to eliminate all the Wilson lines along the themelion. Secondly, the themelion wraps several compact directions, whose radii affect the charge densities and hence the $\alpha_i$. Since these radii typically vary across spacetime, some of the themelion constraints impose conditions on the location of the themelion itself.

\begin{table}
  \renewcommand{\arraystretch}{1.4}
  \centering
  \scriptsize
  \begin{tabular}{|c|>{\centering}p{7ex}|>{\centering}c|p{.5ex}|c|>{\centering}p{7ex}|c|}
  \cline{1-3} \cline{5-7}
  \textbf{Object} & \multicolumn{2}{c|}{\textbf{Coefficient}} & & \textbf{Object} & \multicolumn{2}{c|}{\textbf{Coefficient}} \\
  \cline{1-3} \cline{5-7}
  F1(y) & $\alpha_1$ & \multirow{4}{*}{$\mqty{x_1\\y_1\\z_1}$} & & F1($\psi$) & $\alpha_5$ & \multirow{4}{*}{$\mqty{x_2\\y_2\\z_2}$}\\
  \cline{1-2}\cline{5-6}
  NS5(y1234) & $\alpha_2$ & & & NS5($\psi$1234) & $\alpha_6$ & \\
  \cline{1-2}\cline{5-6}
  P(y) & $\alpha_3$ & & & P($\psi$) & $\alpha_7$ & \\
  \cline{1-2}\cline{5-6}
  KKm(y1234;$\psi$) & $\alpha_4$ & & & KKm($\psi$1234;y) & $\alpha_8$ &\\
  \cline{1-3}
  \cline{5-7}
  D2(y1) & $\alpha_9$ & \multirow{2}{*}{$u_1$} & & D2($\psi$1) & $\alpha_{11}$ & \multirow{2}{*}{$u_2$}\\
  \cline{1-2}\cline{5-6}
  D4(y234) & $\alpha_{10} = -\alpha_9$ & & & D4($\psi$234) & $\alpha_{12} = -\alpha_{11}$ &\\
  \cline{1-3}
  \cline{5-7}
  D0 & $\alpha_{13}$ & \multirow{2}{*}{$v_1$} & & D2(y$\psi$) & $\alpha_{15}$ & \multirow{2}{*}{$v_2$}\\
  \cline{1-2}\cline{5-6}
  D4(1234) & $\alpha_{14} = - \alpha_{13}$ & & & D6(y$\psi$1234) & $\alpha_{16} = -\alpha_{15}$ &\\
  \cline{1-3}
  \cline{5-7}
  F1(1) & $\alpha_{17}$ & \multirow{2}{*}{$w_1$} & & NS5(y$\psi$234) & $\alpha_{19}$ & \multirow{2}{*}{$w_2$}\\
  \cline{1-2}\cline{5-6}
  P(1) & $\alpha_{18} = - \alpha_{17}$ & & & KKm(y$\psi$234; 1) & $\alpha_{20} = -\alpha_{19}$ &\\
  \cline{1-3}
  \cline{5-7}
  \end{tabular}
  \caption{The Type IIA constituents and parameterization of the most general themelion with $\IT^3$ invariance.}
 \vskip-0.5cm
  \label{Table1}
\end{table}

The simplest themelion constraint to interpret is \eqref{mass}, which reflects the fact that the mass density of the themelion is the sum of its global charge densities: $M=Q_1+Q_2+Q_3+Q_4$. The parameters in the second constraint, \eqref{bubble}, depend on distinct powers or the radius of the $\psi$-circle, and so this determines the possible locations of the themelion. In particular, for the bubbling solutions of \cite{Bena:2005va, Berglund:2005vb, Bena:2007kg}, this constraint is equivalent to the bubble equations \footnote{A similar relation between bubble equations and probe-brane constraints has been found for supertubes probing bubbling solutions \cite{Bena:2008dw}}. The last linear equations \eqref{z4charges} reflect the fact that, in the six-dimensional supergravity theories used to build  superstrata and bubbling solutions, the tensor fields corresponding to $\alpha_{9,10,11,12}$ must be anti-self dual.

For bubbling solutions the quadratic constraints, like \eqref{quadratic1}, correspond to smoothness conditions, giving exactly the quadratic constraints on the sources  of harmonic functions needed to construct smooth horizonless solutions of  \cite{Bena:2005va, Berglund:2005vb, Bena:2007kg}. 

As we noted earlier, a superstratum is made of two themelia: one at the supertube locus and one at the center of space. Using the correspondence between parameters in branes in Table \ref{Table1}, one can see that the supertube themelion has $\alpha_{3,4,5,6,11,12}=0$ and $\alpha_7+\alpha_8=\alpha_{9}+\alpha_{10}=0$.  The simplest way to identify the parameters of the center-of-space themelion is to perform a ``spectral inversion,'' $\psi \leftrightarrow y$, \cite{Niehoff:2013kia}  \footnote{We note that  spectral inversion is inconsistent with \eqref{mass} and \eqref{bubble}.  Consistency  can be restored by performing a large gauge transformation in supergravity to eliminate Wilson lines.} and then one sees that this themelion has  $\alpha_{1,2,7,8,9,10}=0$ and $\alpha_{11}+\alpha_{12}= \alpha_{3}+\alpha_{4}=0$. 

Remembering that the supertube themelion has charge densities that depend on $\psi$, and the center-of-space themelion has charge densities that depend on $y$ one sees that  \eqref{quadratic1} imposes two independent constraints on these densities. Amazingly, these are exactly the coiffuring constraints \cite{Bena:2016ypk}, which were necessary to construct a smooth supergravity solutions.

\vspace*{-.5cm}
\section{The \newstratum }
\vspace*{-.3cm}

The themelia that enter in the construction of the superstrata actually belong to a much larger moduli space of themelia. Indeed, the Type IIB superstratum only had fields that preserve the $\IT^4$ invariance, but when we dualize it to the Type IIA/M-Theory duality frame we use in this Letter, it only has a  $\IT^3 $ invariance, along the directions $234$. This suggests  one should consider a more general themelion which preserves this $\IT^3 $ invariance and has branes that can wrap $1, y, z$ and $\psi$. This themelion can have 20 possible species of branes:
\begin{equation}
\begin{aligned}
{\rm M2}(0 a b) \,, \ \ {\rm  M5}(0234ab)  \,, \ \  {\rm P}(a) \,, \ \ {\rm KKM}(0234abc)  \,, 
\end{aligned}
\label{GenThem}
\end{equation}
where $a, b, c \in \{1, y, z,\psi\}$. The complete set of species is described in the IIA nomenclature in Table \ref{Table1}. 

The  projector is now constructed using $\hat P = P+P'$ where $P$ is given by (\ref{Pss1}) and 
\begin{equation}
\begin{aligned}
P' &= \alpha_{13}  \Gamma^{0z}    + \alpha_{14} \Gamma^{01234z}  +  \alpha_{15} \Gamma^{0 y\psi}  + \alpha_{16}   \Gamma^{0y\psi 1234}    \\
&  \  + \alpha_{17}  \Gamma^{01z} +  \alpha_{18}    \Gamma^{01}    +  \alpha_{19} \Gamma^{0y\psi 234}    + \alpha_{20}  \Gamma^{0y\psi234 z} \,.
\end{aligned}
\label{Pss2}
\end{equation}

We find that the null-space condition (\ref{susies2}) now imposes eight linear constraints on the $\alpha_j$, while the projection condition yields another $15$ quadratic constraints.  
This is a hugely overdetermined system, and if we first use the linear constraints, we can re-parameterize the system in terms of three  vectors in $\IC^3$, defined by:
\begin{equation}
\begin{aligned}
\vec p_1 & \equiv  (u_1 + i u_2\,, -(w_1 - i w_2)\,, x_1+ i x_2) \,, \\
\vec p_2 & \equiv (-i(v_1+ i v_2)\,,  -i  (y_1 -i y_2)\,, -i(w_1 - i w_2)) \,, \\
\vec p_3 & \equiv (-(z_1+i z_2)\,,-(v_1+ i v_2)\,,u_1 + i u_2)\,.
\end{aligned}
\label{vecparams}
\end{equation}
where $(u_1, \dots , z_2)$ are real parameters. Note that each vector $p_i$ has an entry in common with both other vectors 
and so there  are twelve independent real parameters and their relationship with the $\alpha_j$ may be found in (\ref{reparam1}). The projection condition, $\hat P^2 =\hat P$, is equivalent to the statement that the $\vec p_j$ are orthonormal  in $\IC^3$  .  This means that one can then use $U(3)$ to rotate the $\vec p_j$ to a simple canonical form:
\begin{equation}
{\vec  p_1}{'}  \equiv  (0\,, 0\,, 1) \,, \  {\vec  p_2}{'}   \equiv (0 \,, \mp i  \,, 0 ) \,, \ {\vec  p_3}{'}     \equiv (\mp 1\,, 0 \,,0 )\,.
\label{vecbase}
\end{equation}
Note that this basis  corresponds to $x_1 =1$, $y_1 = z_1 = \pm 1$, with all the other parameters set to zero.  From  (\ref{reparam1}) one sees that this themelion corresponds to $\alpha_1 =1$ (for $+$),  or   $\alpha_4 =1$  (for $-$) with all the other $\alpha_j$ vanishing.  Thus a $U(3)$  U-duality rotation can {\it locally} map this themelion onto a stack of F1 strings or a stack of coincident KKM's.  

Using the U$(3)$ one can  parameterize the most general themelion, remembering that the rotation must be restricted by the constraints on interrelated components of the $\vec p_i$ in (\ref{vecparams}).  The result is a six parameter family, (\ref{uvwxyz}),  given by the angles, $\theta_1, \theta_2, \varphi_j$, $j=1, 2,3,4$.   
%The angles $\theta_j$ appear in $SO(3) \subset U(3)$ rotations and the $\varphi_j$ introduce complex phases.
{As noted earlier, the  projector of the supertube-locus themelion of the superstratum solution, based  on (\ref{Pss1}), is given by $v_j = w_j =0$, or $\theta_2 = \frac{\pi}{2}$ and $\varphi_2=0$. }

It is not hard to see that our general themelion also include the themelia that enter the construction of the vector superstratum of \cite{Ceplak:2022} of which a subset can be built using only NS-NS fields. The  ``supertube-locus" themelion of these NS-NS vector superstratum is obtained by taking $u_j = v_j =0$, or $\theta_2 = 0$.

To obtain a themelion with no components in the space-time ($\psi$) directions, a glance at  Table \ref{Table1} reveals that one must remove all the constituents in the second column, and hence all the second components must vanish.   This is achieved by setting all the $\varphi$-phases in (\ref{uvwxyz}) to zero.  This leaves a themelion with:
\begin{equation}
\begin{aligned}
P =& (\beta_1  \,  \Gamma^{0yz}  + \beta_2 \,   \Gamma^{01234y}  + \beta_3  \,   \Gamma^{0y} ) +  \beta_4   \,  (   \Gamma^{0z}   - \Gamma^{01234z})\\
&  +  \beta_5\,(   \Gamma^{01}- \Gamma^{01z} )     + \beta_6 \,  ( \Gamma^{01y}  +  \Gamma^{0234yz}  ) \,.
\end{aligned}
%\label{Psl1}
\nonumber
\end{equation}
and 
\begin{equation}
\begin{aligned}
\beta_1 =&  \cos^2 \! \coeff{1}{2} \theta_1 \,, \ \beta_2 = \sin^2  \!  \coeff{1}{2} \theta_1   \sin^2  \!  \theta_2    \,, \ \beta_3 = \sin^2 \! \coeff{1}{2} \theta_1   \cos^2  \! \theta_2  \,,  \\
\beta_4 =&  \sin^2 \! \coeff{1}{2} \theta_1   \sin  \theta_2  \cos  \theta_2\,, \ \beta_5 + i \beta_6 =\coeff{1}{2}  \sin  \theta_1\, e^{i   \theta_2}     \,.
\end{aligned}
%\label{Psl2}
\nonumber
\end{equation}
Amazingly, this is the projector of the super-maze \cite{Bena:2022wpl}.
Hence, the class of themelia we obtain from  $\hat P$ contains the themelia that govern both the original and vector superstrata, as well as the super-maze themelion. Based on the Themelion Conjecture, we expect   there to be supergravity solutions made of multiple generalized themelia, which we can call  {\it \newstrata}.   These will contain all the existing superstrata and super-mazes. Since the momentum charge is carried by different excitations in the superstrata and  the super-maze, we expect the {\it \newstrata} to have a larger entropy than either subclass. 

Moreover, it was argued in \cite{Bena:2022wpl} that, because the super-maze captures brane fractionation,  its entropy is expected to match that of the rotationally-invariant microstates of the black hole, $2 \pi \sqrt{{5 \over 6} Q_1 Q_5 Q_P}$. The {\it \newstratum} will capture  these microstates as well those that break the spacetime rotational invariance of the black-hole horizon. Given that the  \newstratum\ captures fractionation and restores democracy between the $y$-circle and a generic torus direction, we expect it to match the full black-hole entropy. It would be exciting to construct some  {\it \newstratum} solutions, to see whether this intuition is realized.

By breaking the torus invariance, we have found themelia that capture fractionated branes, and perhaps the full black-hole entropy.  It is  thus natural  ask how the phase space of themelia will expand if we relax the  $\IT^3$ invariance imposed here and allow all possible  branes wrapping the compactified dimensions.   Given that  the generic super-maze breaks  this invariance, we suspect that relaxing the $\IT^3$ invariance will lead to a phase space that is a combination of all superstrata and generic supermazes.

Armed with our knowledge of themelia and {\newstrata}, we return to the resolution of a seeming paradox of the original superstratum: its moduli space appears to have a degenerate limit to a BTZ black hole with finite horizon area \cite{Bena:2016ypk,Bena:2017xbt}, and yet a themelion cannot have a horizon.  This degenerate limit  arises when one forces the two themelia of the superstratum to coincide. One of these themelia has only $\psi$ fluctuations and a KKM that shrinks $y$, while the other one has only $y$ fluctuations and a KKM that shrinks $\psi$. Forcing these two themelia to coincide turns off both the $\psi$ and $y$ fluctuations, and either requires one to set the momentum charge to zero, or results in a configuration that is not a themelion.

However, we have seen that the themelia that give rise to the superstratum are part of a much larger family of themelia that can fluctuate not only  along $\psi$ and $y$, but also along $z$ and the torus directions. The \newstratum\ is built  from these more generic themelia, and the coincidence limit is no longer degenerate: the momentum charge can be carried by fluctuations  along $z$ and the torus, and the resulting configuration will be another horizonless themelion,  the super-maze.  The presence of a black hole in the phase space of superstratum solutions is an artifact of objects made from enforcing  $\IT^4$ invariance and {\it smearing} the themelia. This illustrates, once again, the fuzzball precept: horizons only appear because of ensemble averaging.

\vspace{1mm}
\noindent
%%%%%%%%%%%%%%%%%%%%%%%%%%%%%%%%%
{\bf Acknowledgments}.  
%%%%%%%%%%%%%%%%%%%%%%%%%%%%%%%%%
%\vspace{-2mm}
This work  is supported in part by the ERC Grants 787320 - QBH Structure and 772408 - Stringlandscape.
The work of NPW is supported in part by the DOE grant DE-SC0011687.

%%%%%%%%%%%%%%%%%%%%%%%%%%%%%%%%%%%%%
\appendix
%
%%%%%%%%%%%%%%%%%%%%%%%%%%%%%%%%%%%%%
\section{Parameterizing the themelia}
\label{Sect:Appendix}
%%%%%%%%%%%%%%%%%%%%%%%%%%%%%%%%%%%%%

In the interest of efficiency we parameterize the most general themelion considered in this paper, which should be the building block of   \newstrata,   via:
\begin{equation}
\begin{aligned}
\alpha_1 & =  \coeff{1}{4}\, (1+x_1+y_1 +z_1) \,, \    \alpha_2  =  \coeff{1}{4}\, (1-x_1+y_1 -z_1) \,, \\
 \alpha_3 & =  \coeff{1}{4}\, (1-x_1-y_1 + z_1) \,, \  \alpha_4  =  \coeff{1}{4}\, (1+x_1-y_1 -z_1) \,, \\
 \alpha_5 & =  \coeff{1}{4}\, (x_2+y_2 +z_2) \,, \    \alpha_6  =  \coeff{1}{4}\, (-x_2+ y_2 - z_2) \,, \\
 \alpha_7 & =  \coeff{1}{4}\, (-x_2-y_2 + z_2) \,, \  \alpha_8 =  \coeff{1}{4}\, (x_2-y_2 - z_2) \,, \\
 \alpha_9 & \!=\! \alpha_{10}  \!=\!   \coeff{1}{2}\,u_1   \,, \,   \alpha_{11}  \!=\!   \alpha_{12}  \!=\!   \coeff{1}{2}\, u_2\,, \,     \alpha_{13}  \!=\!   -\alpha_{14} \!=\!   \coeff{1}{2}\,v_1 \,, \\
 \alpha_{15} &  \!=\!  -\alpha_{16}  \!=\!   \coeff{1}{2}\,v_2  ,     	\alpha_{17}  \!=\!  - \alpha_{18}  \!=\!  \coeff{1}{2}\,w_1 ,      \alpha_{19}  \!=\!   -\alpha_{20} \!=\!  \coeff{1}{2}\,w_2   .
\end{aligned}
\label{reparam1}
\end{equation}
and the solution to the projection conditions is:
\begin{equation}
\begin{aligned}
&u_1 +  i u_2 = s_1 s_2 \,  e^{i \varphi_1}  \,, \\
&v_1 +  i v_2 = s_2 c_2 \, e^{i (\varphi_1-\varphi_2-\varphi_3)} (e^{-2i \varphi_4} - c_1) \,,      \\
& w_1 +  i w_2 = s_1 c_2 \, e^{i \varphi_2} \,,  \quad x_1 +  i x_2 = c_1 e^{i \varphi_3}  \,, \quad \\
&y_1 +  i y_2 = e^{i (2\varphi_2 + \varphi_3)} \, ( c_1 c_2^2 + s_2^2 \,e^{-2i \varphi_4}  ) \,,      \\
&z_1 +  i z_2 = e^{i (2\varphi_1 -\varphi_3)}\,  ( c_2^2 \, e^{2i \varphi_4} + c_1 s_2^2 ) \,,
 \end{aligned}
\label{uvwxyz}
\end{equation}
%l
where   $c_j  \equiv \cos \theta_j$ and $s_j \equiv  \sin \theta_j$.  The quadratic terms appear because some $U(3)$ rotation angles must be fixed in terms of others to preserve the relationships between the components, (\ref{vecparams}) of the $\vec p_j$.

\bibliography{microstates}

\end{document}